\documentclass[british]{jfm}
\usepackage[T1]{fontenc}
\usepackage[utf8]{inputenc}
\usepackage{amsmath}
\usepackage{amssymb}
\usepackage{graphicx}
\usepackage{esint}
\usepackage[authoryear]{natbib}
\usepackage[hyperindex,breaklinks]{hyperref}

\makeatletter

\providecommand{\tabularnewline}{\\}
\renewcommand\[{\begin{equation}} 
\renewcommand\]{\end{equation}}
\renewenvironment{align*}{\align}{\endalign}

\makeatother

\usepackage{babel}
\begin{document}
\global\long\def\dt#1{\frac{\mathrm{d}#1}{\mathrm{d}t}}

\global\long\def\Dt#1{\frac{\mathrm{D}#1}{\mathrm{D}t}}

\global\long\def\pp#1#2{\frac{\partial#1}{\partial#2}}

\global\long\def\v#1{\boldsymbol{#1}}

\global\long\def\mm#1{\mathrm{#1}}
\author{Anton Daitche} 
\affiliation{Institute for Theoretical Physics, Westf\"alische Wilhelms-Universit\"at,\\ Wilhelm-Klemm-Str. 9, 48149 M\"unster, Germany\\ anton.d@wwu.de}

\title{On the role of the history force for inertial particles in turbulence}
\maketitle
\begin{abstract}
The history force is one of the hydrodynamic forces which act on a
particle moving through a fluid. It is an integral over the full time
history of the particle's motion and significantly complicates the
equations of motion (accordingly it is often neglected). We present
here a study of the influence of this force on particles moving in
a turbulent flow, for a wide range of particle parameters. It is shown
that the magnitude of history force can be significant and that it
can have a considerable effect on the particles' slip velocity, acceleration,
preferential concentration and collision rate. We also investigate
the parameter dependence of the strength of these effects.
\end{abstract}

\section{Introduction\label{sec:Introduction}}

The advection of finite-size particles by a fluid flow plays an important
role in a large number of natural and industrial situations, a few
examples are: cloud microphysics, combustion processes and formation
of marine aggregates or planets. The underlying flow is turbulent
in many cases. When the inertia of the particles is small enough they
can be accurately modelled as \emph{tracer}s, i.e. particles which
follow the fluid flow exactly. When this is not the case the particles
can deviate from the flow due to their inertia and one then speaks
of \emph{inertial particles.} A fundamental difference between these
two types of particles is that the dynamics of tracers preserves phase
space volume (when the flow is incompressible) whereas the dynamics
of inertial particles is dissipative, thus allowing for the existence
of attractors. As a direct consequence, inertial particles can accumulate
in certain regions of the flow -- an effect termed preferential concentration
-- while tracers stay homogeneously distributed for all times. 

The equations of motion for a particle in a fluid flow can (under
certain assumptions) be systematically derived from the Navier-Stokes
equation, as has been done by \citet{Maxey1983,Gatignol1983} (see
the references therein for a historical overview of the literature
before 1983). These equations of motion are comprised of different
terms which makes it possible to decompose the total hydrodynamical
force on the particle into different contributions like the Stokes
drag, the added mass effect and the history force. The latter is an
integral over the whole history of the particle's motion -- a memory
effect. \citet{Boussinesq1885} and \citet{Basset1888} were the first
to point out the existence of the history force, which is therefore
often called the Boussinesq-Basset force. 

Before we proceed to the full equation, let us present an intuitive
argument for the existence of the history force and try to guess its
form. Consider a spherical particle moving through a still fluid with
a constant velocity. If the Reynolds number is small, we can use the
linearised Navier-Stokes equation and the flow is then given by the
Stokes flow around a sphere. Now imagine a sudden change $\mm d\v v_{0}$
in the velocity of the particle at time $t=0$. This disturbance will
lead to the generation of vorticity at the particle's surface, which
will decay diffusively and thus as $1/\sqrt{t}$ in the vicinity of
the particle. We might expect that the force on the particle due to
this decaying part of the flow will be proportional to $\mm d\v v_{0}/\sqrt{t}$%
\footnote{There is actually another effect of the disturbance $\mm d\v v_{0}$:
the added mass effect, which is proportional to $\mm d\v v_{0}/\mm dt$
and describes the immediate pressure response of the fluid. We neglect
it here for the sake of a concise illustration of the history force.%
}. This is already a memory effect because the particle ``remembers''
the disturbance at $t=0$ in all following instants. The particle
obviously does not have a memory itself but rather the information
is stored in the flow around it. Now imagine a general motion of the
particle as a series of velocity jumps $\mm d\v v(\tau)$. Then the
superposition of the flows due to these disturbances will produce
a force proportional to
\begin{equation}
\int_{0}^{t}\frac{1}{\sqrt{t-\tau}}\mm d\v v(\tau)=\int_{0}^{t}\frac{1}{\sqrt{t-\tau}}\frac{\mm d\v v}{\mm d\tau}\mm d\tau\label{eq:naive-history-force}
\end{equation}
(a superposition is admissible because of the small Reynolds number
and thus the approximate linearity of the Navier-Stokes equation).
Up to a constant prefactor (\ref{eq:naive-history-force}) is the
history force on a particle moving in a still fluid. It describes
the effect of the decaying parts of the flow which were generated
by the acceleration of the particle at previous instants (on the particle
at the present instant).

In the case of a general fluid flow the motion of a spherical particle
is described by the following equation
\begin{equation}
\dt{\v v}=\beta\Dt{\v u}-\frac{1}{\tau_{p}}\left(\v v-\v u\right)-\sqrt{\frac{3\beta}{\tau_{p}}}\frac{1}{\sqrt{\pi}}\dt{}\int_{0}^{t}\frac{\v v-\v u}{\sqrt{t-\tau}}\mm d\tau.\label{eq:MR-dimensionless}
\end{equation}
This is the version derived by \citet{Maxey1983,Gatignol1983} with
the slightly different (and widely used) form of the added mass term
by \citet{Auton1988}. Here $\v v\equiv\mm d\v x/\mm dt$, $\v u=\v u(\v x,t)$
and $\Dt{\v u}=\partial_{t}\v u+\v u\cdot\nabla\v u$ denote the particle
velocity, the fluid velocity and the fluid acceleration at the position
of the particle; $\v u$ is the undisturbed flow, without the particle's
presence. There are two parameters in (\ref{eq:MR-dimensionless}),
the first is the density parameter 
\begin{equation}
\beta=\frac{3}{2\varrho+1},\label{eq:beta}
\end{equation}
where $\varrho=\varrho_{p}/\varrho_{f}$ is the particle's density
$\varrho_{p}$ normalized by that of the fluid $\varrho_{f}$. When
$\varrho>1$ the particle is heavier than the fluid and $0<\beta<1$;
when $\varrho<1$ it is lighter than the fluid and $1<\beta<3$. The
second parameter is the particle response time
\begin{equation}
\tau_{p}=\frac{1}{3\beta}\frac{r^{2}}{\nu},\label{eq:tp-definition}
\end{equation}
where $r$ is the radius of the particle and $\nu$ the kinematic
viscosity of the fluid. We have omitted here the so-called Faxén corrections
and the influence of gravity. The terms appearing on the right-hand
side of (\ref{eq:MR-dimensionless}) are the pressure gradient (which
contains a contribution from the added mass term), the Stokes drag
and the history force. The latter is an integral over the whole history
of the particle and describes the effect of the decaying disturbance
flow generated by the particle at earlier times on the particle at
the present time, in harmony with the intuitive argument described
above. The history force is a viscous effect and is sometimes referred
to as the unsteady drag. 

In turbulence the relevant time scale to which the particle response
time $\tau_{p}$ should be compared is the Kolmogorov-time $\tau_{\eta}=\sqrt{\nu/\epsilon}$,
where $\epsilon$ is the mean energy dissipation. The Stokes number
\begin{equation}
St=\frac{\tau_{p}}{\tau_{\eta}}\label{eq:St}
\end{equation}
is the ratio of these two characteristic time scales and is a measure
for the importance of the particle's inertia. Throughout the paper
we will use the Kolmogorov scales of turbulence $\eta=\left(\nu^{3}/\epsilon\right)^{1/4}$
(length), $\tau_{\eta}$ (time), $u_{\eta}=\eta/\tau_{\eta}$ (velocity)
and $a_{\eta}=u_{\eta}/\tau_{\eta}$ (acceleration) to normalize dimensional
quantities. These scales are the smallest scales of the turbulent
motion and thus relevant for small particles, which ``live'' on
the smallest scales, e.g. for the particles considered here $\tau_{p}/\tau_{\eta}$
and $r/\eta$ are at most on the order of a few times unity.

Note that in (\ref{eq:MR-dimensionless}) the form of the history
force is non-standard: the derivative is outside the integral. This
form is actually more general as it is valid for initial conditions
with $\v v(0)\neq\v u(0)$. The following relation (derived via partial
integration) shows the connection between the generalized form and
the standard form
\begin{equation}
\dt{}\int_{0}^{t}\frac{\v v-\v u}{\sqrt{t-\tau}}\mm d\tau=\int_{0}^{t}\frac{\frac{\mathrm{d}}{\mathrm{d}\tau}\left(\v v-\v u\right)}{\sqrt{t-\tau}}\,\mathrm{d}\tau+\frac{\v v(0)-\v u(0)}{\sqrt{t}}=\sqrt{\pi}\left(\dt{}\right)^{1/2}\left(\v v-\v u\right).\label{eq:generelized-history-force}
\end{equation}
The term $\left(\v v(0)-\v u(0)\right)/\sqrt{t}$ should be added
in (\ref{eq:MR-dimensionless}) when $\v v(0)\neq\v u(0)$ as pointed
out by \citet{Michaelides1992,Maxey1993}. Thus the left-hand side
in (\ref{eq:generelized-history-force}) is a form of the history
force which is valid for any type of initial condition. This generalized
form is also exactly the definition of the fractional derivative of
Riemann-Liouville type of order $1/2$ (see, e.g. \citet{Podlubny1998}).
This connection to fractional derivatives and relation (\ref{eq:generelized-history-force})
was first pointed out by \citet{Tatom1988}. We will use the fractional
derivative notation to abbreviate the history force in the following.

The history force has been frequently neglected in applications of
the equation of motion (\ref{eq:MR-dimensionless}). This is often
done to simplify the problem: the history force turns the equation
of motion into an integro-differential equation and thus makes it
much more difficult to reason about the solutions. For example, the
existence, uniqueness, regularity and asympotitics of the solutions
of (\ref{eq:MR-dimensionless}) have been only recently studied by
\citet{Farazmand2015,Langlois2014}. Another difficulty is the computation
of numerical solutions with the history force: on the one hand there
is the singularity of the integrand of the history force which impedes
an accurate numerical approximation and on the other hand there is
the necessity to recompute the history integral for every new time
step which causes high numerical costs. The first problem can be resolved
by an appropriate treatment of the history force and, by now, higher-order
integration schemes are available (\citet{Daitche2013}). The second
problem is inherent to the dynamics with memory and can be attenuated
only by an approximation of the history kernel (see e.g. \citet{Hinsber2011}).

The effect of the history force in non-chaotic flows has been studied
analytically by \citet{DO1994,Coimbra1998,Hill2005,Angilella2004,Candelier2014,Lim2014}
and experimentally by \citet{Mordant2000,Abbad2004,Coimbra2004,Lohse2006,Lohse2009}.
It was shown that memory can be quite important, for example the experimental
studies showed that the history force is necessary (in these cases)
for a match between experiment and theory. Chaotic dynamics of inertial
particles have been studied by \citet{YRK1997,Daitche2011,Guseva2013,Daitche2014}
showing that the history force can \emph{qualitatively }change the
dynamics, for example, the history force reduces the tendency for
accumulation and can change the nature of attractors from non-chaotic
to chaotic and vice versa. \citet{Reeks1984,Mei1991,Aartrijk2010}
have shown that memory can affect the dispersion of particles in turbulence.
\citet{Calzavarini2012} studied the influence of Faxén corrections,
non-linear drag and history force on the statistical properties of
large neutrally buoyant particles ($r/\eta\in\left[3,30\right]$)
moving in a turbulent flow and found a significant influence of all
three terms (especially for particles considerably larger than $\eta$).
The importance of different forces acting on a particle in turbulence
has been studied by \citet{Armenio2001,Olivieri2014} showing that
the magnitude of the history force can be significant and that it
can also alter the contribution of the other forces. \citet{Olivieri2014}
also found that memory can reduce preferential concentration. We will
present a detailed comparisons with the results by \citet{Olivieri2014}
in section~\ref{sec:Comparison-with-Olivieri}, where we find a considerable
difference to our results.

One of the conditions for the derivation of (\ref{eq:MR-dimensionless})
is that the particle Reynolds number $Re_{p}=r\left|\v v-\v u\right|/\nu$
is small. \citet{Lovalenti1993} and \citet{Mei1994} have developed
extended versions of (\ref{eq:MR-dimensionless}) for finite $Re_{p}$
(up to a few hundreds by \citet{Mei1994}). Both versions contain
significantly more complicated forms of the history force; Mei's version
also includes a nonlinear form of the drag. \citet{Maxey1996} found
that up to $Re_{p}\approx17$ the standard form (\ref{eq:MR-dimensionless})
``may be quite adequate in practice even though not justified by
theory''. In an experimental study \citet{Abbad2004} considered
particle Reynolds numbers of up to $0.5$ and found the standard form
to remain valid.  In our simulations the typical $Re_{p}$ is smaller
or on the order of unity (except for $\varrho=0$ and $St>2$, when
it becomes larger than $3$). We thus assume that the standard version
(\ref{eq:MR-dimensionless}) is applicable in our case. As we want
to concentrate on the effects of the history force we do not consider
here any further corrections, like e.g. the lift force. In a comparison
of the magnitudes of the Faxén corrections and the history force we
found that the former are \emph{much} smaller in most cases considered
here, see appendix~\ref{sec:faxen-corrections}. Thus we neglect
the Faxén corrections in the following.

The aim of the present paper is to provide a comprehensive study of
the role of the history force for the motion of particles in a turbulent
flow. To this end we prepared numerical simulations for a wide range
of particle parameters, which will be investigated for effects of
memory on different statistical particle properties. An important
question we want to answer here is for which particle parameters the
effects of memory are important.

The paper is structured as follows: The next section will present
details on the numerical simulations. The magnitude of the history
force will be studied in section~\ref{sec:Forces}, followed by investigations
of its effect on the slip velocity $\v v-\v u$ (section~\ref{sec:Slip-velocity}),
the acceleration (section~\ref{sec:Acceleration}), preferential
concentration (section~\ref{sec:Preferential-concentration}) and
collision rates (section~\ref{sec:Collision-rates}). Section~\ref{sec:Comparison-with-Olivieri}
will detail a comparison with the recent work by \citet{Olivieri2014},
followed by a summary and discussion in section~\ref{sec:Discussion-and-conclusion}.
Appendix~\ref{sec:faxen-corrections} presents a comparison of the
history force and the Faxén corrections. A note on the case of neutrally
buoyant particles, which is not considered in the main part of text,
is given in appendix~\ref{sec:Neutrally-buoyant-particles}. The
numerical scheme for the integration of particle trajectories with
the history force and the method of forcing the turbulence are described
in the appendices \ref{sec:Time-stepping-scheme-appendix} and \ref{sec:Forcing},
respectively.

\section{Numerical simulations}

\begin{table}
\begin{centering}
\begin{tabular}{cccccccccc}
$Re_{\lambda}$ & $L_{{\rm box}}/\eta$ & $L/\eta$ & $\lambda/\eta$ & $\Delta x/\eta$ & $T_{\mm{sim}}/\tau_{\eta}$ & $T/\tau_{\eta}$ & $\Delta t/\tau_{\eta}$ & $u_{\mm{rms}}/u_{\eta}$ & $N^{3}$\tabularnewline
\hline 
113 & 632 & 157 & 20.9 & 1.23 & 997 & 29.1 & 0.0199 & 5.40 & $512^{3}$\tabularnewline
\end{tabular}
\par\end{centering}

\protect\caption{\label{tab:sim-para}Parameters of the simulated turbulent flow: Taylor
Reynolds number $Re_{\lambda}=\lambda u_{\text{rms}}/\nu$, size of
the periodic box $L_{{\rm box}}$, integral scale $L=u_{\protect\mm{rms}}^{3}/\epsilon$,
Taylor microscale $\lambda=u_{\protect\mm{rms}}\sqrt{15\nu/\epsilon}$,
size of a grid cell $\Delta x$, length of the whole simulation $T_{\protect\mm{sim}}$,
large-eddy turnover time $T=L/u_{\protect\mm{rms}}$, time step $\Delta t$,
root-mean-square of the velocity $u_{\protect\mm{rms}}=\sqrt{\left\langle \protect\v u^{2}\right\rangle /3}$,
number of grid points $N^{3}$. All dimensional quantities are given
in multiples of the corresponding Kolmogorov scales.}
\end{table}

The turbulent flow is generated in a triply periodic box by a large
scale forcing (as described in appendix~\ref{sec:Forcing}). We solve
the vorticity equation, which is equivalent to the incompressible
Naiver-Stokes equation, by a standard dealiased Fourier-pseudo-spectral
method (\citet{Canuto1987,Hou2007}) with a third-order Runge-Kutta
time-stepping scheme (\citet{shu88jcp}). The particle trajectories
are integrated with a specialized scheme which treats the history
force appropriately; it is a modified version of the third-order scheme
developed by \citet{Daitche2013} and is described in appendix~\ref{sec:Time-stepping-scheme-appendix}.
The values of Eulerian quantities, which are present on a grid, are
obtained at the particle positions through tricubic interpolation.
The turbulent flow is statistically homogeneous, isotropic and stationary,
with the flow characteristics depicted in table~\ref{tab:sim-para}. 

For this study a number of simulations with different particle parameters
have been prepared. For the density $\varrho$ the values $1000$,
$100$, $10$, $2$, $0.5$, $0$ were chosen. $\varrho=1000$ corresponds
to water droplets moving in air, relevant for, e.g., cloud microphysics.
Density ratios $10$ and $2$ are close to those of metals and sand
in water. $\varrho=1$ is the case of neutrally buoyant particles,
which is not discussed in the main text but in appendix~\ref{sec:Neutrally-buoyant-particles}.
$\varrho=0$ corresponds to air bubbles in water. The Stokes number
has been varied in the range $\left[0.1,3.0\right]$. It is typically
in this range where interesting properties of inertial particles appear.
Also, this range is constrained from above by the limitations of (\ref{eq:MR-dimensionless})
(the particle Reynolds number grows with $St$) and from below by
the time step of the simulation (it should be significantly smaller
than $\tau_{p}$). For every combination of the values of $\varrho$
and $St$ a simulation of particles with and without the history force
has been prepared. The number of simulated particles for every parameter
combination is $N_{p}=10^{5}$, except for the investigations in sections
\ref{sec:Preferential-concentration} and \ref{sec:Collision-rates}
where $N_{p}=5\cdot10^{5}$ (in this case the length of the simulations
is $T_{\mm{sim}}=700\tau_{\eta}=24T$). The initial particle positions
have been chosen randomly and homogeneously distributed in space;
the initial particle velocity is that of the fluid at the particle's
position. Statistical quantities (e.g. averages) have been obtained
by sampling over the particle ensemble and over time. To assure that
the particles have equilibrated with the flow, this sampling is started
after an initial period of five large-eddy turnover times.

\section{Forces\label{sec:Forces}}

We start by comparing the magnitudes of the different forces acting
on a particle. In the following, the moduli of the particle acceleration
and of the three terms on the right-hand side of (\ref{eq:MR-dimensionless})
will be denoted by $a$, $a_{P}$ (pressure gradient), $a_{S}$ (Stokes
drag) and $a_{H}$ (history force), respectively. 

\begin{figure}
\begin{centering}
\includegraphics{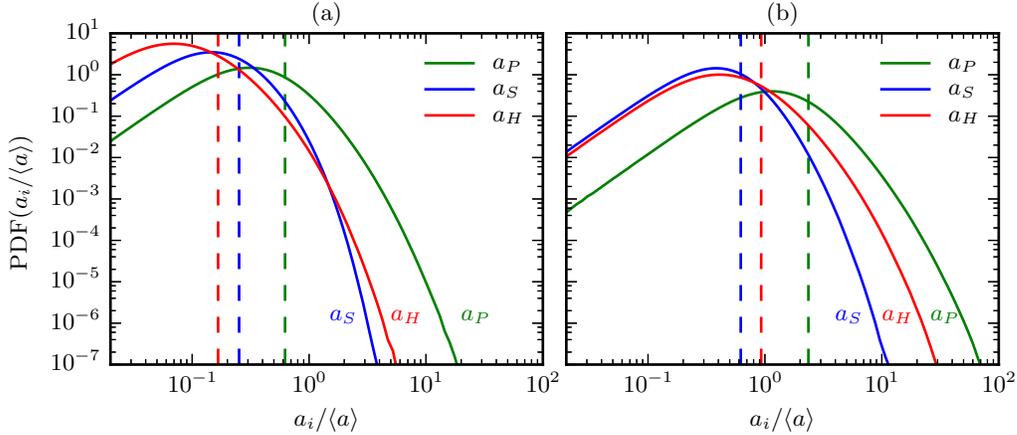}
\par\end{centering}

\protect\caption{\label{fig:force-pdfs} PDFs of the different accelerations $a_{i}$
due to the pressure gradient ($i=P$), the Stokes drag ($i=S$) and
the history force ($i=H$). The particle parameters are (a) $\varrho=2$,
$St=0.5$ ($r/\eta=0.95$) and (b) $\varrho=0$, $St=0.5$ ($r/\eta=2.1$)
. The vertical lines show the positions of the averages: $\left\langle a_{P}\right\rangle /\left\langle a\right\rangle =0.62$,
$\left\langle a_{S}\right\rangle /\left\langle a\right\rangle =0.25$,
$\left\langle a_{H}\right\rangle /\left\langle a\right\rangle =0.17$
in (a) and $\left\langle a_{P}\right\rangle /\left\langle a\right\rangle =2.4$,
$\left\langle a_{S}\right\rangle /\left\langle a\right\rangle =0.62$,
$\left\langle a_{H}\right\rangle /\left\langle a\right\rangle =0.93$
in (b).}
\end{figure}

Figures \ref{fig:force-pdfs}a and \ref{fig:force-pdfs}b show the
probability density functions (PDFs) of $a_{P}$, $a_{S}$ and $a_{H}$
for particles heavier ($\varrho=2$) and lighter ($\varrho=0$) than
the fluid. In both cases $a_{P}$ is the dominant contribution, followed
by $a_{S}$ and $a_{H}$. The Stokes drag and the history force are
of similar magnitude showing that the history force is a viscous effect
which can be as important as the drag. In these particular cases we
find $\left\langle a_{H}\right\rangle /\left\langle a_{S}\right\rangle =66\%$
in figure~\ref{fig:force-pdfs}a and $\left\langle a_{H}\right\rangle /\left\langle a_{S}\right\rangle =150\%$
in figure~\ref{fig:force-pdfs}b. Additionally the PDF of the history
force has longer tails than that of the Stokes drag (see figure~\ref{fig:force-pdfs}a
and also note that the $x$-axis is logarithmic), i.e. the history
force has more extreme events.

Let us now consider the \emph{relative }importance of the history
force and compare it to the Stokes drag by means of the ratio $a_{H}/a_{S}$.
Before proceeding to the numerical simulations, let us try to obtain
an estimate of this ratio. To this end we apply the approximation
\begin{equation}
\left(\dt{}\right)^{1/2}\left(\v v-\v u\right)\approx\frac{\alpha}{\sqrt{\tau_{\eta}}}\left(\v v-\v u\right),\label{eq:history-force-approximation}
\end{equation}
which is motivated by the fact that the history force is a (fractional)
time-derivative. The choice of $\tau_{\eta}$ as the characteristic
time scale of $\v v-\v u$ seems natural as the particle ``lives''
on the small scales and should follow the flow to some degree (except
for very large $St$). The constant $\alpha$ is expected to be on
the order of unity. With this approximation and the definition of
$a_{H}$ and $a_{S}$ we obtain an estimate for the relative magnitude
of the history force (see also \citet{Daitche2014}):
\begin{equation}
\frac{a_{H}}{a_{S}}\approx\alpha\sqrt{\frac{3\beta\tau_{p}}{\tau_{\eta}}}=\alpha\frac{r}{\eta}.\label{eq:HS-ratio}
\end{equation}
Thus we expect the importance of the history force to scale solely
with the particle size and to be independent of its density.

\begin{figure}
\begin{centering}
\includegraphics{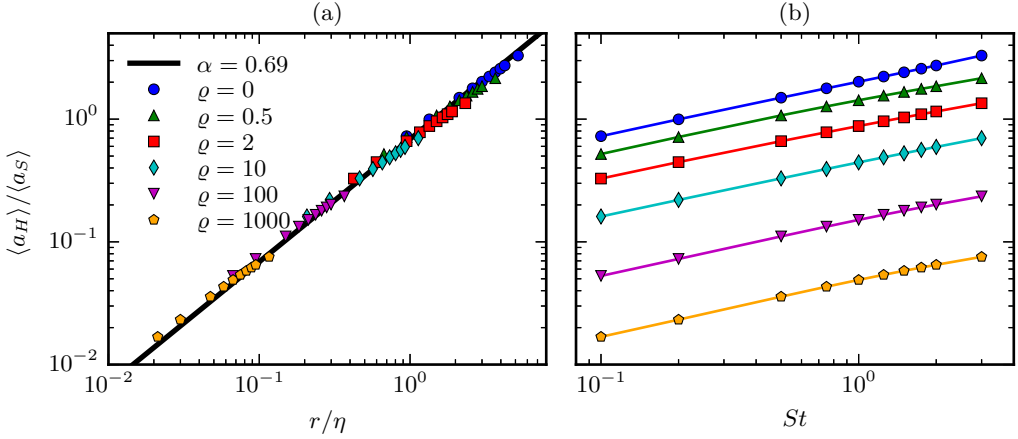}
\par\end{centering}

\protect\caption{\label{fig:HS-ratio}The ratio of the average magnitude of the history
force and the Stokes drag as a function of (a) $r/\eta$ and (b) $St$.}
\end{figure}

Figure \ref{fig:HS-ratio}a shows $\left\langle a_{H}\right\rangle /\left\langle a_{S}\right\rangle $,
measured in the turbulence simulations, as a function of $r/\eta$.
We see a collapse of the data for different densities, which confirms
our expectation that the particle size is the determining parameter
for the relative magnitude of the history force. (When looking more
closely we actually see a very weak dependence on the density, the
collapse is not perfect.) A fit of our prediction (\ref{eq:HS-ratio})
to the data yields $\alpha=0.69$ and we see that this simple estimate
holds quite well. 

Figure~\ref{fig:HS-ratio}b shows $\left\langle a_{H}\right\rangle /\left\langle a_{S}\right\rangle $
as a function of the Stokes number. In this case there is no collapse
for different densities and the relative magnitude of the history
force decays with growing density. This tendency is true for a \emph{fixed}
Stokes number. Because $St$ depends on both the density and the particle
size, increasing $\varrho$ and holding $St$ fixed means that we
implicitly decrease the particle size. Thus the dependence on $\varrho$
in figure~\ref{fig:HS-ratio}b is actually a disguised dependence
on $r$. That there is no ``real'' dependence on $\varrho$ can
be concluded from figure~\ref{fig:HS-ratio}a.

We are now able to make an informed categorization of the importance
of the history force based on its relative magnitude. For $r/\eta$
around $10^{-2}$ or less its contribution is very small (less than
1\%) and should be negligible in most cases. For $r/\eta$ around
$0.1$ the contribution is around 7\%; while not negligible the history
force is expected to play a minor role. For $r$ on the order of $\eta$
the contribution of the history force is on the order of the Stokes
drag and is thus expected to be important for the particle motion.

\section{Slip velocity\label{sec:Slip-velocity}}

\begin{figure}
\begin{centering}
\includegraphics{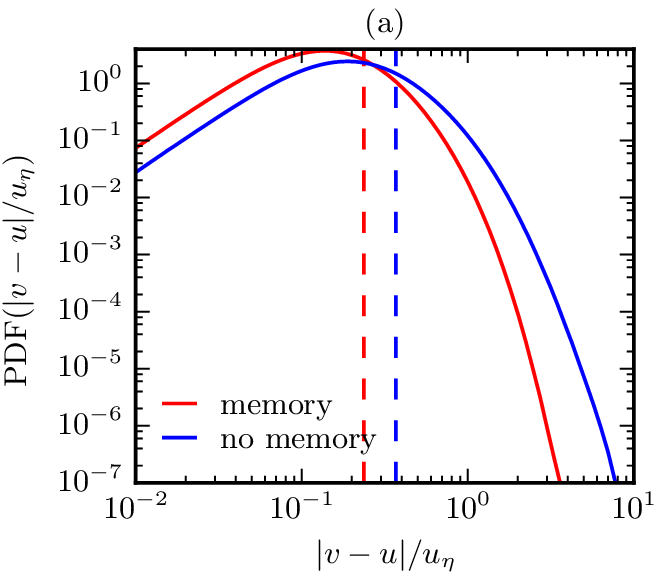}\includegraphics{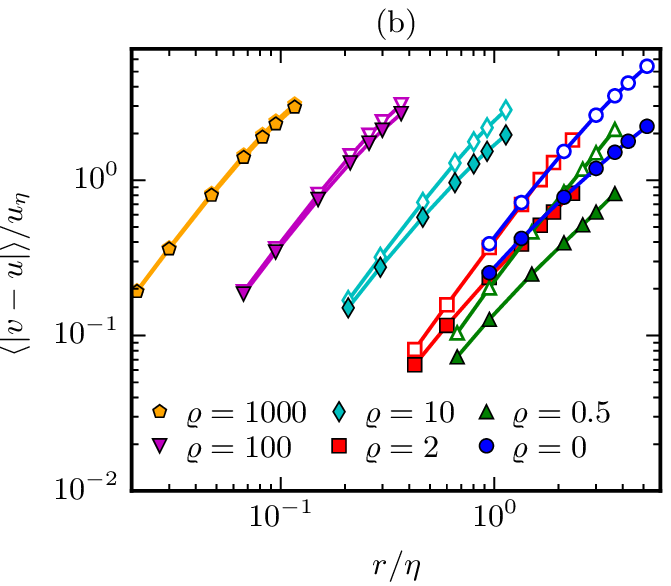}
\par\end{centering}

\protect\caption{\label{fig:slipvelocity}(a) PDFs of the slip velocity for particles
with and without memory for $\varrho=2$ and $St=0.5$ ($r/\eta=0.95$).
The vertical lines show the position of the averages: $\left\langle \left|\protect\v v-\protect\v u\right|/u_{\eta}\right\rangle _{\text{memory}}=0.24$
and $\left\langle \left|\protect\v v-\protect\v u\right|/u_{\eta}\right\rangle _{\text{no memory}}=0.37$.
(b) The average slip velocity with (filled symbols) and without memory
(unfilled symbols) as a function of the particle size and for different
particle densities.}
\end{figure}

Now that we can estimate the magnitude of the history force, let us
turn to its effects on the particle dynamics. A basic property of
inertial particles is that they can deviate from the fluid flow (in
contrast to tracers). It is thus natural to study the effect of the
history force on this deviation, namely on the slip velocity $\v v-\v u$.
Figure~\ref{fig:slipvelocity}a shows the PDF of $\left|\v v-\v u\right|$
with and without the history force (denoted by ``memory'' and ``no
memory'' respectively). Memory leads to a reduction of the slip velocity;
in this case the mean slip velocity is reduced by 35\% in comparison
to the case without memory. Also, the tails of the PDF become shorter,
i.e., strong detachments from the flow occur less frequently with
memory. Figure~\ref{fig:slipvelocity}b presents an overview of the
full parameter space, showing the average slip velocity with and without
memory. The natural parameter for the slip velocity is the Stokes
number%
\footnote{More precisely, the weakly inertial limit (\ref{eq:weakly-inertial-limit})
suggests that it is proportional to $St\left|\beta-1\right|$. %
}, however to make the role of the size clearer the data are shown
as a function of $r/\eta$. We see that the effect of the history
force (the difference between the case with and without memory) increases
with $r$. 

\begin{figure}
\begin{centering}
\includegraphics{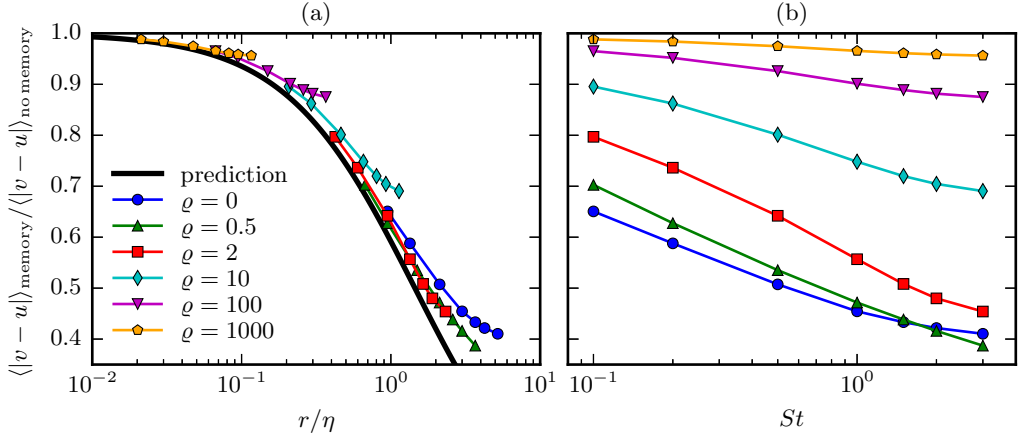}
\par\end{centering}

\protect\caption{\label{fig:slipvelocity-ratio}The ratio of the average slip velocity
with and without memory as a function of (a) $r/\eta$ and (b) $St$.}
\end{figure}

To obtain a compact representation of this effect of memory let us
use the ratio $\left|\v v-\v u\right|_{\mm{memory}}/\left|\v v-\v u\right|_{\mm{no\, memory}}$,
i.e. the slip velocity with the history force divided by the one without.
Before proceeding to the data from the simulations let us first try
to obtain an analytical estimate. Using the equation of motion (\ref{eq:MR-dimensionless})
and the approximation (\ref{eq:history-force-approximation}) we can
estimate the slip velocity as

\begin{equation}
\v v-\v u\approx\left(\beta\Dt{\v u}-\dt{\v v}\right)\left(\frac{1}{\tau_{p}}+\alpha\sqrt{\frac{3\beta}{\tau_{p}\tau_{\eta}}}\right)^{-1}=\left(\beta\Dt{\v u}-\dt{\v v}\right)\frac{\tau_{p}}{1+\alpha\frac{r}{\eta}},\label{eq:slip-velocity-approximation}
\end{equation}
where the case without memory is contained for $\alpha=0$. If we
assume that the first factor on the right-hand side of (\ref{eq:slip-velocity-approximation})
is the same for both cases we obtain
\begin{equation}
\frac{\left|\v v-\v u\right|_{\mm{memory}}}{\left|\v v-\v u\right|_{\mm{no\, memory}}}\approx\frac{1}{1+\alpha\frac{r}{\eta}}.\label{eq:slip-velocity-formula}
\end{equation}
Parameter $\alpha$ was introduced in section~\ref{sec:Forces},
where its value was determined from the numerical simulations: $\alpha=0.69$.
Thus (\ref{eq:slip-velocity-formula}) provides a prediction for the
slip velocity ratio. Figure~\ref{fig:slipvelocity-ratio}a shows
the ratio $\left\langle \left|\v v-\v u\right|\right\rangle _{\mm{memory}}/\left\langle \left|\v v-\v u\right|\right\rangle _{\mm{no\, memory}}$
obtained from the simulations along with this prediction. Although
not perfect, this estimate -- based on rather simple assumptions --
fits reasonably well to the simulations (note that the black line
in figure~\ref{fig:slipvelocity-ratio}a is not a fit of (\ref{eq:slip-velocity-formula})
to the data; we used the previously obtained value of $\alpha$).
An important observation is that the slip velocity ratio collapses
for different densities when considered as a function of $r$. Although
this collapse is not perfect (there seems to be a systematic deviation
for $St>1$), it is still remarkable that the density plays only a
minor role also for the effect of memory on the slip velocity. When
the slip velocity ratio is considered as a function of $St$, there
is no collapse at all, see figure~\ref{fig:slipvelocity-ratio}b.

This result suggests again that the natural parameter for the effect
of the history force is the particle size. Due to its simplicity,
formula (\ref{eq:slip-velocity-formula}), as well as (\ref{eq:HS-ratio}),
can be very useful for estimates of this effect (when $\alpha$ is
not known, one might assume $\alpha\approx1$).

The findings of this section support the categorization of the memory
effect for different particle sizes given at the end of the previous
section. For $r/\eta\leq10^{-2}$ the effect is negligible as the
relative reduction of the slip velocity ( $1-\left\langle \left|\v v-\v u\right|\right\rangle _{\mm{memory}}/\left\langle \left|\v v-\v u\right|\right\rangle _{\mm{no\, memory}}$)
is below 1\%. For $r/\eta\approx0.1$ the reduction is around 5\%,
i.e. not completely negligible but still small. For $r/\eta\approx1$
it is around 35\% and thus considerable.

\section{Acceleration\label{sec:Acceleration}}

\begin{figure}
\begin{centering}
\includegraphics{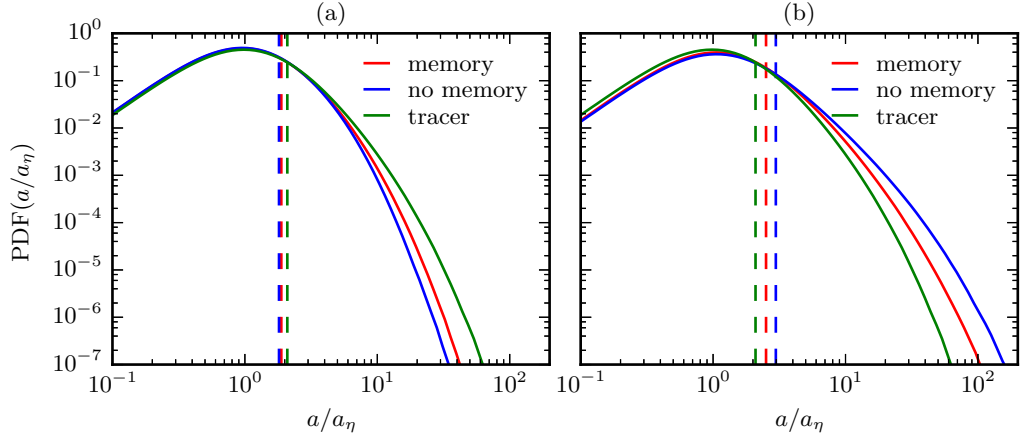}
\par\end{centering}

\protect\caption{\label{fig:acceleration-pdfs}PDFs of the magnitude of the acceleration
for particles with and without memory and tracers. The particle parameters
are (a) $\varrho=2$, $St=0.5$ ($r/\eta=0.95$) and (b) $\varrho=0$,
$St=0.5$ ($r/\eta=2.1$). The vertical lines show the positions of
the averages: $\left\langle a/a_{\eta}\right\rangle _{\text{tracer}}=2.1$
and $\left\langle a/a_{\eta}\right\rangle _{\text{memory}}=1.9$,
$\left\langle a/a_{\eta}\right\rangle _{\text{no memory}}=1.8$ in
(a) and $\left\langle a/a_{\eta}\right\rangle _{\text{memory}}=2.5$,
$\left\langle a/a_{\eta}\right\rangle _{\text{no memory}}=3.0$ in
(b).}
\end{figure}

Let us now turn to the influence of the history force on the particle
acceleration, a much studied quantity in turbulence research. Figures
\ref{fig:acceleration-pdfs}a and \ref{fig:acceleration-pdfs}b show
the PDFs of the modulus of the acceleration for particles with and
without memory and tracers. For heavy particles (figure~\ref{fig:acceleration-pdfs}a)
the history force \emph{increases} the acceleration: the average increases
slightly and the tails more notably (note the logarithmic $x$-axis).
For light particles (figure~\ref{fig:acceleration-pdfs}b) the effect
is opposite, the history force \emph{decreases} the acceleration.
The change of the mean acceleration is stronger in this case. We can
describe these observations in a unified way: the acceleration PDF
with memory comes closer to that of tracers (compare figures \ref{fig:acceleration-pdfs}a
and \ref{fig:acceleration-pdfs}b).

\begin{figure}
\begin{centering}
\includegraphics{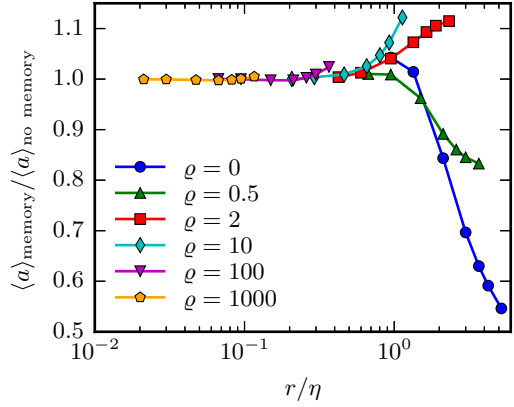}
\par\end{centering}

\protect\caption{\label{fig:ratio-accel}The ratio of the averaged acceleration magnitudes
with and without memory.}
\end{figure}

To obtain an overview of the parameter dependence we again turn to
ratios. Figure~\ref{fig:ratio-accel} shows the ratio of the (averaged)
magnitudes of acceleration with and without memory. We see that the
history force generally increases the acceleration of heavy particles
and decreases the acceleration of light particles, as we found in
two particular cases above. This effect is noticeable but not very
large, except for $\varrho=0$ and large particle sizes. Note that
there is no collapse of the ratios shown in figure~\ref{fig:ratio-accel}
in contrast to the slip velocity ratio shown in figure~\ref{fig:slipvelocity-ratio}a
and the force ratio shown in figure~\ref{fig:HS-ratio}a. Although
the influence of memory generally increases with $r$ also for this
observable, there is additionally a clear dependence on the density.

\section{Preferential concentration\label{sec:Preferential-concentration}}

\begin{figure}
\begin{centering}
\includegraphics{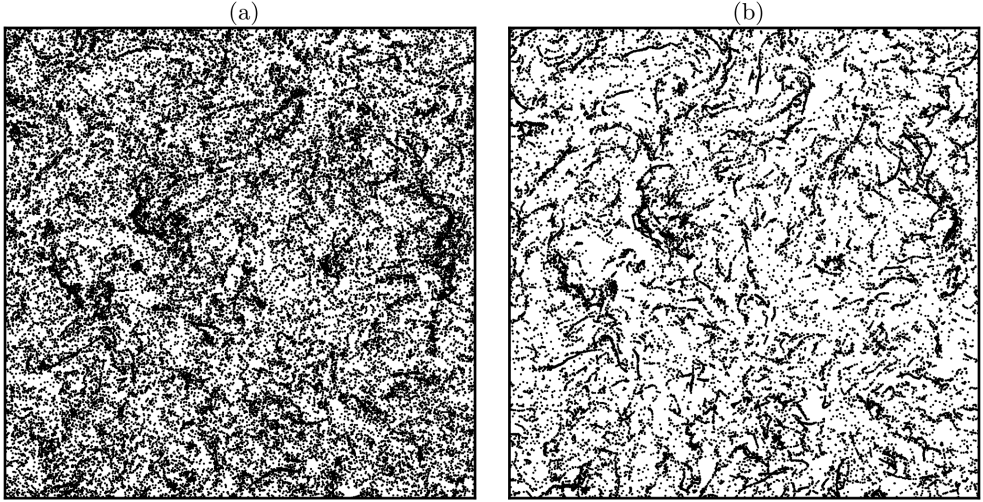}
\par\end{centering}

\protect\caption{\label{fig:particle-distribution}The distribution of light particles
(a) with and (b) without memory. Both plots show the positions of
$5\cdot10^{4}$ particles contained in the slab $\left[0,L_{\protect\mm{box}}\right]\times\left[0,L_{\protect\mm{box}}\right]\times\left[0,L_{\protect\mm{box}}/10\right]$
at $t=291\tau_{\eta}$ (this corresponds to 10 large-eddy-turnover
times). The particles were homogeneously distributed at the beginning
of the simulation and have the parameters $\varrho=0$ and $St=0.5$. }
\end{figure}

A much studied phenomenon concerning inertial particles is preferential
concentration: the tendency of particles to gather in certain regions
of the flow. Figure~\ref{fig:particle-distribution} shows the distribution
of light particles with and without memory. In both cases we can see
an inhomogeneous distribution of particles, i.e. preferential concentration.
We also clearly see that the preferential concentration is \emph{weaker}
with the history force (figures \ref{fig:particle-distribution}a
and \ref{fig:particle-distribution}b show the same number of particles).

To obtain an overview of the parameter dependence of this effect we
need to quantify preferential concentration. There are several possible
approaches; we use the correlation dimension $D_{2}$ as introduced
by \citet{Grassberger1983}. Let $P(R)$ be the probability to find
a given particle pair with a distance below $R$, i.e.
\[
P(R)=\left\langle \Theta\left(R-\left|\v x_{1}-\v x_{2}\right|\right)\right\rangle ,
\]
where $\Theta$ is the Heaviside function and $\v x_{1}$, $\v x_{2}$
the positions of the two particles. Then the correlation dimension
is defined through
\[
P(R)\propto R^{D_{2}}\quad R\rightarrow0.
\]
In studies of preferential concentration the radial distribution function
$g(R)$ is frequently used. It is closely related to $P(R)$ and $D_{2}$
($d$ is the space dimension): 
\[
g(R)\propto\frac{\partial_{R}P(R)}{R^{d-1}}\propto R^{D_{2}-d},
\]
see e.g. \citet{Bec2005}. Thus $D_{2}$ can also be seen as a characterization
of the radial distribution function. 

\begin{figure}
\includegraphics{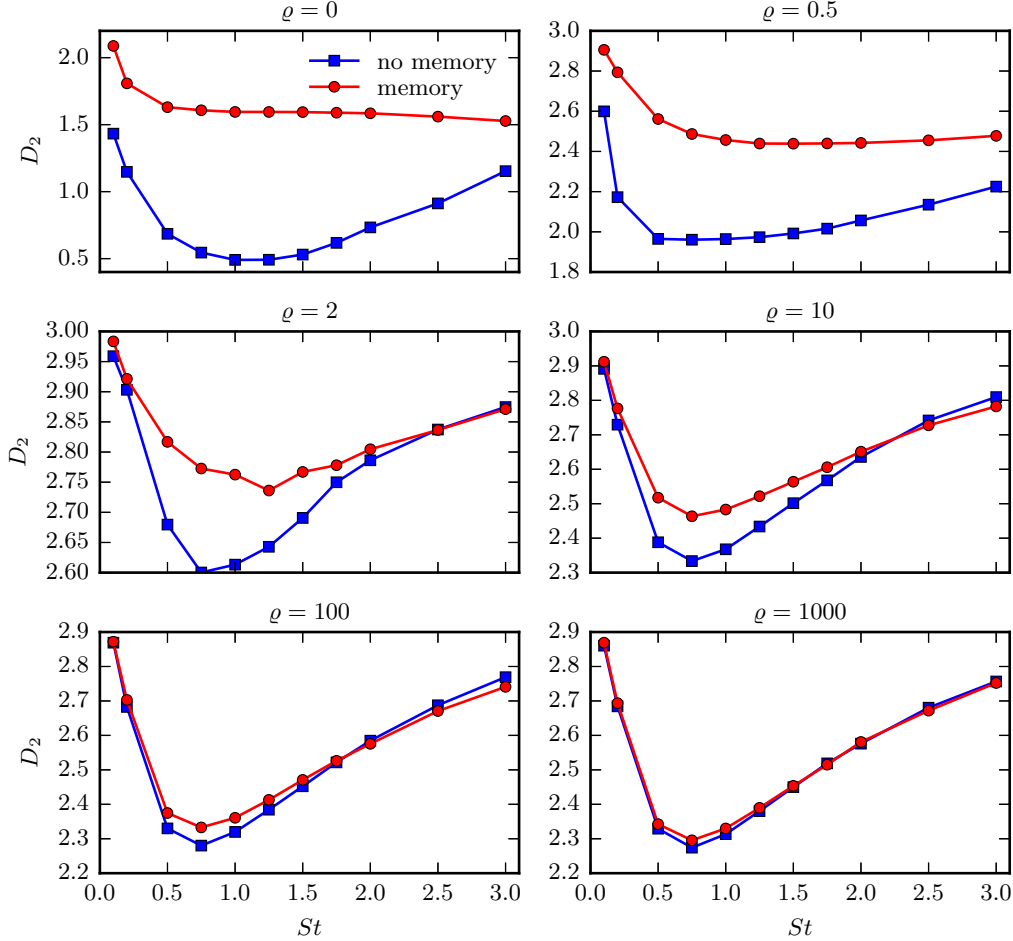}

\protect\caption{\label{fig:D2}The correlation dimension $D_{2}$ of the particle
distribution in space as a function of $St$ for different densities
with and without memory.}
\end{figure}

To determine $D_{2}$ we measured $P(R)$ in our simulations and fitted
a power law in the range $R/\eta\in\left[10^{-1},10\right]$. Figure~\ref{fig:D2}
shows the correlation dimension as a function of the Stokes number
for different densities. Without memory we see the typical unimodal
shape of $D_{2}(St)$ found in previous studies, e.g. \citet{Calzavarini2008}.
The minimal value of $D_{2}$, corresponding to maximal preferential
concentration, is obtained around $St\approx1$. Concerning the influence
of the history force we find: for $\varrho=1000$ there is hardly
any effect; for $\varrho=100$ there is a small effect; for $\varrho=2$
and $\varrho=10$ the effect is quite noticeable; for light particles
($\varrho=0$ and $\varrho=0.5$) the effect is strong, e.g. for $\varrho=0$
the smallest value of $D_{2}$ increases by $1.1$ when memory is
included -- a very considerable change for a fractal dimension. It
is interesting to note that the position of the smallest $D_{2}$
is shifted by the history force for $\varrho=2$; for $\varrho=0.5$
and $\varrho=0$ there does not seem to be a clear minimum with memory
but rather an extended plateau. Also note that the influence of memory
is strongest where $D_{2}$ attains its minimum. To sum up, if we
fix the Stokes number and decrease $\varrho$, the effect of the history
force on $D_{2}$ and thus on preferential concentration increases
and becomes very strong for small densities. 

\begin{figure}
\begin{centering}
\includegraphics{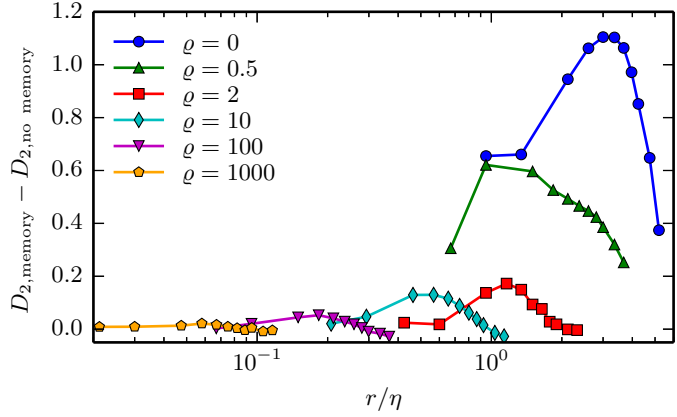}
\par\end{centering}

\protect\caption{\label{fig:D2-r-rho}The change in the correlation dimension $D_{2}$
due to memory as a function of the particle size and density.}
\end{figure}

Note that for a \emph{fixed} Stokes number we cannot discern the dependencies
of this effect on $r$ and $\varrho.$ As $St$ depends on both parameters,
varying $\varrho$ implicitly varies $r$; in particular, decreasing
$\varrho$ increases $r$, see (\ref{eq:beta})-(\ref{eq:St}). Thus
the dependence on $\varrho$ in figure~\ref{fig:D2} might be a disguised
dependence on $r$ (as is the case in figure~\ref{fig:HS-ratio}b
for the relative magnitude of the history force and in figure~\ref{fig:slipvelocity-ratio}b
for the slip velocity ratio). Thus we have to consider this effect
as a function of the particle size and density to discern the dependencies.
Figure~\ref{fig:D2-r-rho} shows the change of the correlation dimension
$D_{2,\mm{memory}}-D_{2,\mm{no\ memory}}$ as a function of these
two variables. We see a dependence on \emph{both} $r$ and $\varrho$.
This is in contrast to the behaviour of the relative magnitude of
the history force (section~\ref{sec:Forces}) and the slip velocity
ratio (section~\ref{sec:Slip-velocity}), where the determining parameter
was solely $r$ and similar to the case of the acceleration (previous
section). Note also that the dependence on $r$ and $\varrho$ is
quite non-trivial, e.g. it is non-monotonic in both variables.

A good way to summarize these findings is as follows: For a fixed
Stokes number the influence of the history force increases \emph{monotonically}
when the particle density decreases or (equivalently) when the particle
size increases. However, one has to keep in mind that this dependency
cannot be attributed to only one of the two parameters.

\section{Collision rates\label{sec:Collision-rates}}

The collision rate of particles in turbulent flows is important for
the understanding of, e.g., cloud microphysics or aggregation processes;
it is also closely related to preferential concentration. One of the
main findings in this area is that turbulence can enhance the collision
rate enormously. We will study here the effect of the history force
on the collision rate.

Let $N_{c}(t,\Delta\tau)$ be the number of particle collisions during
a small time interval $\left[t,t+\Delta\tau\right]$, then the collision
rate per unit volume is 
\[
R(t)=\frac{N_{c}(t,\Delta\tau)}{L_{\mm{box}}^{3}\Delta\tau}.
\]
This quantity is independent of $t$ in stationary turbulence, but
grows quadratically with the particle number density $n=N_{p}/L_{\mm{box}}^{3}$.
A quantity where this dependence is factored out is the collision
kernel 
\[
\Gamma=\frac{2R}{n^{2}},
\]
which we will use in the following.

To obtain $N_{c}(t,\Delta\tau)$ we have simulated an ensemble of
$N_{p}=5\cdot10^{5}$ particles for every parameter value and saved
snapshots of the particle positions every $\Delta\tau=0.2\tau_{\eta}$
time units. The collisions have been determined using the so-called
ghost collision approximation, i.e. in our simulations the particles
do not interact on contact but simply slip through each other. A collision
is considered to occur when the distance between two particles becomes
less then $2r$. To determine the occurrence of such events, the particle
trajectories have been interpolated linearly in between the saved
snapshots.

\begin{figure}
\includegraphics{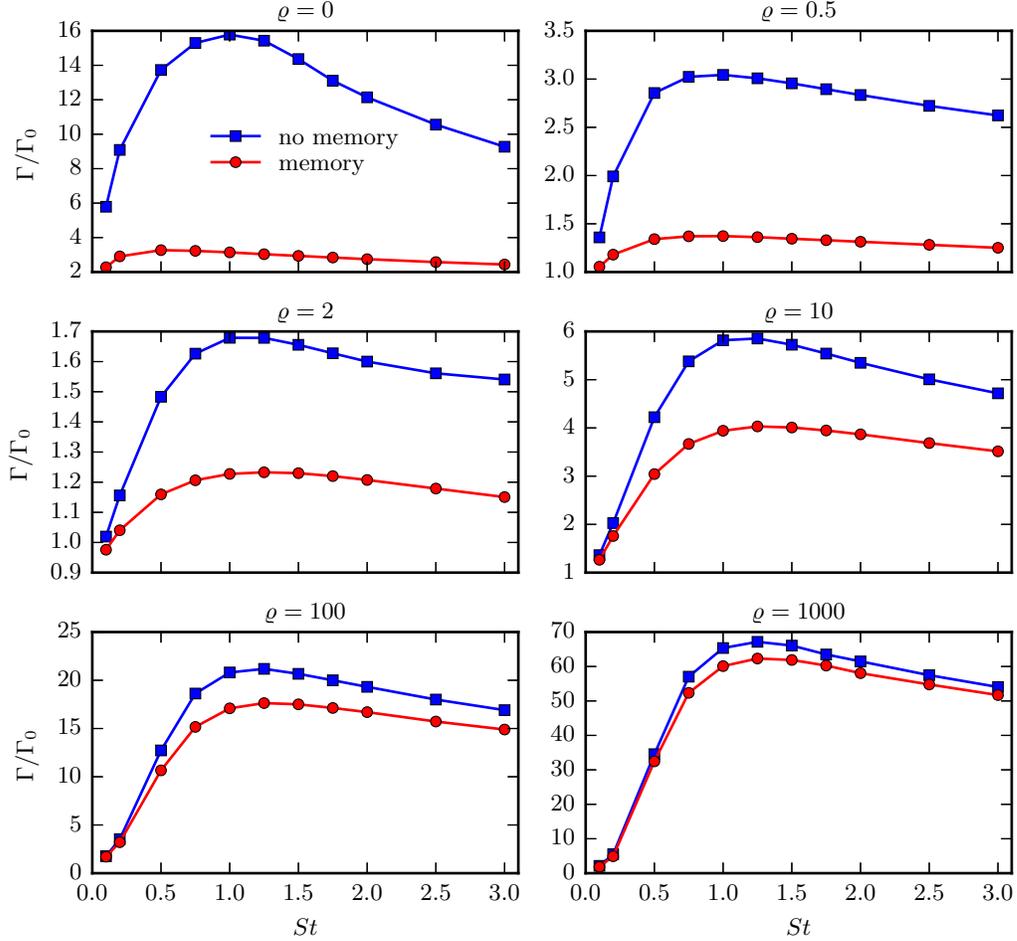}

\protect\caption{\label{fig:collision-kernel}The collision kernel $\Gamma$ as a function
of the Stokes number for different densities with and without memory.
It is normalized by $\Gamma_{0}=\sqrt{8\pi/15}(2r)^{3}/\tau_{\eta}$
the collision kernel based on the dynamics of tracers, as derived
by \citet{Saffman1956}.}
\end{figure}

Figure~\ref{fig:collision-kernel} shows the collision kernel as
a function of the Stokes number for different densities. We see that
the inclusion of the history force leads to a \emph{decrease} in the
collision rate. This effect is very strong for small densities and
decreases for denser particles (when $St$ is fixed). The general
picture is quite similar to that of the previous section where preferential
concentration was studied and we see that memory can be important
in determining the collision rate of inertial particles.

An interesting difference to the case of preferential concentration
is that for large $\varrho$ the influence of memory on collision
rates is stronger than on preferential concentration (compare figures
\ref{fig:D2} and \ref{fig:collision-kernel}). For example at $\varrho=1000$
and $St=1$ there is a very small change in $D_{2}$ due to memory
but a more noticeable one in $\Gamma$ (8\% relative change). The
collision rate in turbulence is generally believed to be influenced
by two effects: preferential concentration and caustics (also called
sling effect), see e.g. \citet{Vosskuhle2014}. Therefore this larger
increase in the collision rate, as compared to preferential concentration,
might be due to an influence of the history force on caustics; or
more specifically, on the relative velocity of colliding particles.

As detailed in the last section, we are not able to discern the roles
of $r$ and $\varrho$ when $St$ is fixed, as is the case in figure~\ref{fig:collision-kernel}.
This can be only done by looking at the change of $\Gamma$ as a function
of $r$ and $\varrho$, as has been done for the correlation dimension
in figure~\ref{fig:D2-r-rho}. In the analogous figure for $\Gamma_{\mm{memory}}/\Gamma_{\mm{no\, memory}}$
(not shown here) one observes a similar picture: the influence of
memory depends on \emph{both} $r$ and $\varrho$. The summary of
the findings for collision rates is similar as well: for a fixed Stokes
number the influence of memory increases monotonically when the particles
density decreases or (equivalently) when the particle size increases.

\section{Comparison with the work of \citet{Olivieri2014}\label{sec:Comparison-with-Olivieri}}

Recently \citet{Olivieri2014} published a study of the history force
in turbulence. During the preparation of the present work discrepancies
with this publication have been found. The type of flow used by \citet{Olivieri2014}
is very similar to ours, therefore we present here a detailed quantitative
comparison. We note that there is a difference in the methods of forcing
the turbulence, which might be a cause of the discrepancies. Our forcing
(which is described in appendix~\ref{sec:Forcing}) varies on a \emph{large}
time scale\emph{ }while the forcing by \citet{Olivieri2014} changes
randomly at every time step (L. Brandt, private communication, 2014),
i.e. the flow and the particles are forced on \emph{small} time scales.

\citet{Olivieri2014} analyse the contributions of different forces
to the whole acceleration of the particle. In our notation their quantities
are

\begin{align}
\v a_{\textsc{pg}} & =\frac{1}{\varrho}\Dt{\v u}\label{eq:a-PG}\\
\v a_{\textsc{am}} & =\frac{1}{2\varrho}\left(\Dt{\v u}-\dt{\v v}\right)\label{eq:a-AM}\\
\v a_{\textsc{sd}} & =\xi\frac{\v u-\v v}{\tau_{p}}\label{eq:a-SD}\\
\v a_{\textsc{ba}} & =\xi\sqrt{\frac{3\beta}{\tau_{p}}}\left(\dt{}\right)^{1/2}\left(\v u-\v v\right)\label{eq:a-BA}
\end{align}
representing the pressure gradient, the added mass, the Stokes drag
and the (Basset) history force. The factor $\xi=\tau_{p}/\tau_{p}'=\left(2\varrho+1\right)/(2\varrho)$
accounts for the slightly different definition of the particle response
time by \citet{Olivieri2014}: $\tau_{p}'=2\varrho r^{2}/(9\nu)$.
We will use the corresponding Stokes number $St'=\tau_{p}'/\tau_{\eta}=St/\xi$
for the particle parameters in this section. Note that $\xi$ and
$\beta$ are shorthand notations for certain combinations of $\varrho$
and that these three quantities are equivalent to each other. 

To quantify the contributions of the different forces \citet{Olivieri2014}
consider the component-wise ratios of the accelerations given above
and the full particle acceleration $\v a\equiv\mm d\v v/\mm dt=\v a_{\textsc{pg}}+\v a_{\textsc{am}}+\v a_{\textsc{sd}}+\v a_{\textsc{ba}}$:
\begin{equation}
\frac{a_{\textsc{pg},x}}{a_{x}},\,\frac{a_{\textsc{sd},x}}{a_{x}},\,\frac{a_{\textsc{ba},x}}{a_{x}}.\label{eq:olivieri-ratios}
\end{equation}
Because of the statistical isotropy of the underlying turbulent flow,
it suffices to consider only one component (we choose the $x$-component).
We omit here the contribution of the added mass effect as it is fully
determined by pressure gradient contribution: $a_{\textsc{am},x}/a_{x}=\frac{1}{2}\left(a_{\textsc{pg},x}/a_{x}+1/\varrho\right)$. 

\begin{figure}
\begin{centering}
\includegraphics{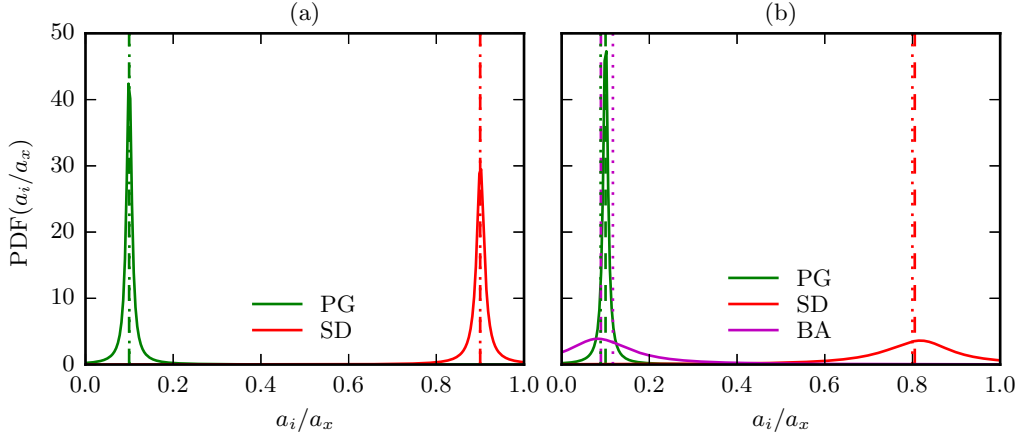}
\par\end{centering}

\protect\caption{\label{fig:olivieri-rho10}PDFs of the ratios  $a_{i,x}/a_{x}$, $i\in\left\{ \text{PG, SD, BA}\right\} $
for $\varrho=10$ and $St'=0.1$ without (a) and with (b) memory.
The vertical dashed lines show the positions of the averages of the
PDFs. The average values predicted by the approximate formulas (\ref{eq:prediction-PG})-(\ref{eq:prediction-BA})
are shown with vertical dotted lines. }
\end{figure}
Figure~\ref{fig:olivieri-rho10} shows the PDFs of the three ratios
(\ref{eq:olivieri-ratios}) for $\varrho=10$, $St'=0.1$ in our simulations
with and without memory. (Note that the choice of colours in the figures
of this section is different from rest of the paper; it is the same
as in (\citet{Olivieri2014}), to facilitate the comparison). Let
us consider the case without memory first. It is shown in figure~\ref{fig:olivieri-rho10}a
and corresponds to figure~$3a$ in (\citet{Olivieri2014}). When
comparing those figures one sees a considerable difference. First,
the mean values (represented by dashed lines) are different, e.g.
in our simulations we find that the Stokes drag on average contributes
to 90\% of the particle acceleration while \citet{Olivieri2014} find
75\%. Second, the PDFs obtained from our simulations are much sharper
and well separated while in (\citet{Olivieri2014}) they are rather
broad and overlap considerably. The case with memory is similar, compare
our figure~\ref{fig:olivieri-rho10}b and figure~$3c$ in (\citet{Olivieri2014}).
The mean value of $a_{\textsc{ba},x}/a_{x}$ (representing the contribution
of the history force) found by \citet{Olivieri2014} is similar to
the Stokes drag's contribution while we find it to be much smaller.
Note that the differences appear also for the case without memory
and thus cannot stem solely from the treatment of the history force.

Let us get an independent estimate of the ratios (\ref{eq:olivieri-ratios}).
To this end we use the approximation (\ref{eq:history-force-approximation})
and the weakly inertial estimate for the slip velocity
\begin{equation}
\v u-\v v\approx\tau_{p}\left(1-\beta\right)\Dt{\v u}\label{eq:weakly-inertial-limit}
\end{equation}
(which can be obtained by expanding (\ref{eq:MR-dimensionless}) in
powers of $\tau_{p}$ and keeping the linear terms, see e.g. \citet{DO1994})
and obtain estimates for (\ref{eq:a-SD}), (\ref{eq:a-BA}) and the
total acceleration
\begin{align}
\v a_{\textsc{sd}} & \approx\xi\left(1-\beta\right)\Dt{\v u}\\
\v a_{\textsc{ba}} & \approx\xi\alpha\frac{r}{\eta}\left(1-\beta\right)\Dt{\v u}\\
\v a & \approx\left(1+\alpha\frac{r}{\eta}\left(1-\beta\right)\right)\Dt{\v u}.\label{eq:a-estimate}
\end{align}
The case without memory is contained for $\alpha=0$. For the ratios
(\ref{eq:olivieri-ratios}) we get estimates which depend on the particle
parameters and $\alpha$ only:
\begin{align}
\frac{a_{\textsc{pg},x}}{a_{x}} & \approx\frac{2}{3}\frac{\xi\beta}{1+\frac{r}{\eta}\alpha\left(1-\beta\right)}\label{eq:prediction-PG}\\
\frac{a_{\textsc{sd},x}}{a_{x}} & \approx\frac{\xi\left(1-\beta\right)}{1+\alpha\frac{r}{\eta}\left(1-\beta\right)}\\
\frac{a_{\textsc{ba},x}}{a_{x}} & \approx\frac{\xi\alpha\frac{r}{\eta}\left(1-\beta\right)}{1+\alpha\frac{r}{\eta}\left(1-\beta\right)}\label{eq:prediction-BA}
\end{align}
These estimates are shown in figure~\ref{fig:olivieri-rho10} as
dotted lines. For the case without memory they coincide perfectly
with the mean values obtained from the simulations (the dashed and
dotted lines overlap). With memory the agreement between the estimate
and the measured values is not perfect but still good. The main source
of error here is probably the value of the coefficient $\alpha$.
We thus see that the above estimates fit quite well to our simulations
but not to the ones by \citet{Olivieri2014}.

\begin{figure}
\begin{centering}
\includegraphics{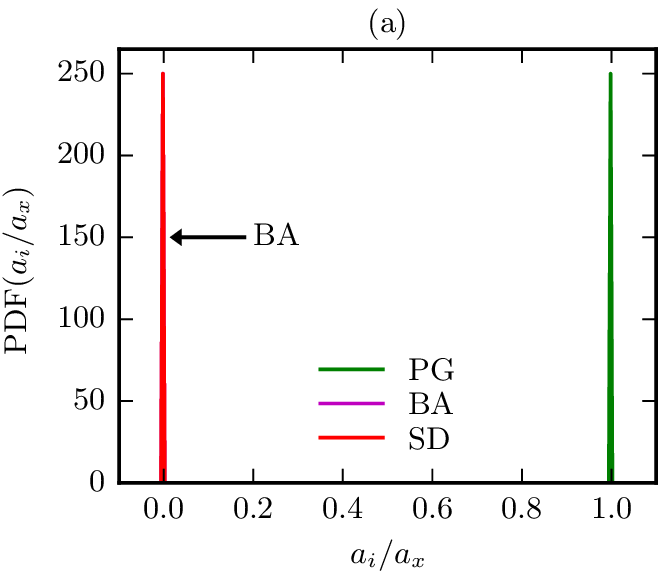}~~\includegraphics{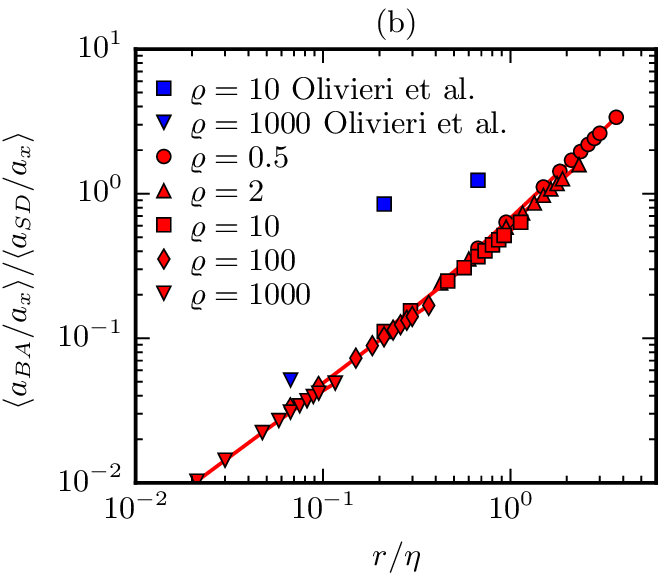}
\par\end{centering}

\protect\caption{\label{fig:olivieri-rho1}(a) PDFs of the ratios $a_{i,x}/a_{x}$
$i\in\left\{ \text{PG, SD, BA}\right\} $ for $\varrho=1$ and $St'=0.01$
with memory. (b) The ratio (\ref{eq:olivieri-BA-SD-ratio}) as a function
of the particle size from our simulations and those by \citet{Olivieri2014}.}
\end{figure}

Let us come to the case of neutrally buoyant particles. Figure~\ref{fig:olivieri-rho1}a
shows the PDFs of the ratios (\ref{eq:olivieri-ratios}) for $\varrho=1$
and $St'=0.01$ with memory (without memory the picture is the same).
The PDFs are sharp peaks around the values $0$ and $1$. We find
that the contribution by the Stokes drag and the history force are
negligible and basically the full contribution to the acceleration
is from the pressure gradient $\v a_{\textsc{pg}}$. This is expected
for neutrally buoyant particles with very small Stokes numbers ($St'=0.01$
here) as they should behave like tracers: with $\v v=\v u$ we get
$\v a_{\textsc{sd}}=\v a_{\textsc{ba}}=0$, $\v a=\v a_{\textsc{pg}}=\mm D\v u/\mm Dt$
and thus $a_{\textsc{sd},x}/a_{x}=a_{\textsc{ba},x}/a_{x}=0$, $a_{\textsc{pg},x}/a_{x}=1$.
This picture is very different in (\citet{Olivieri2014}), see their
figure~4. The average contribution of the Stokes drag is far from
zero and even larger than the contribution of the pressure gradient.
Another strong difference is that the PDFs are rather broad in contrast
to the sharp peaks found here.

To obtain an overview of the deviations (for $\varrho\neq1$) let
us consider the ratio 
\begin{equation}
\frac{\left\langle a_{\textsc{ba},x}/a_{x}\right\rangle }{\left\langle a_{\textsc{sd},x}/a_{x}\right\rangle }\label{eq:olivieri-BA-SD-ratio}
\end{equation}
with the aim to describe the importance of the history force relative
to the Stokes drag, akin to $\left\langle a_{H}\right\rangle /\left\langle a_{S}\right\rangle $
studied in section~\ref{sec:Forces}. Figure~\ref{fig:olivieri-rho1}b
shows the ratio (\ref{eq:olivieri-BA-SD-ratio}) obtained from our
simulations (the dependence on $r/\eta$ is similar to that in figure~\ref{fig:HS-ratio}a)
and compares these with three data points obtained from (\citet{Olivieri2014}).
We see that the difference is considerable, the largest deviation
is by almost one order of magnitude (for $\varrho=10$, $St'=0.1$
and thus $r/\eta=0.21$).

To quantify preferential concentration \citet{Olivieri2014} use the
radial distribution function $g(R)$ and find that memory reduces
$g(R)$ and thus preferential concentration. Generally this is what
we also find, but the amount of reduction is much smaller in our case.
For example, at $St'=1$ we find the relative reduction 
\[
1-\frac{g(R=r/\eta)_{\mm{memory}}}{g(R=r/\eta)_{\mm{no\, memory}}}
\]
to be 25\% for $\varrho=10$ and 3.5\% for $\varrho=1000$ while \citet{Olivieri2014}
find 52\% and 23\%, respectively.

To conclude, we see that there are  strong differences between our
results and those by \citet{Olivieri2014}. A comparison with analytical
estimates of the forces favours our results. From our perspective
the simulations by \citet{Olivieri2014} seem to overestimate the
importance of the history force. We note however that the conclusions
by \citet{Olivieri2014} are similar to ours on a more general, qualitative
level.

\section{Summary and discussion\label{sec:Discussion-and-conclusion}}

In the present paper we we have studied the importance of the the
history force -- a memory effect -- for the motion of inertial particles
in a turbulent flow. We have found that it can be quite important
(depending on the particle parameters). Its effect is to \emph{reduce}
(i) the slip velocity, (ii) the difference in acceleration between
inertial particles and tracers, (iii) the preferential concentration
and (iv) the collision rate. We can concisely summarize these finds
as follows: the history force causes inertial particles to stay closer
to the flow and to behave more like tracers. This is in accordance
with previous findings in a smooth flow with chaotic advection (\citet{Daitche2011,Daitche2014}).

How large the effects of memory are depends on the particle parameters.
We have found simple approximative relations for the magnitude of
the history force relative to the Stokes drag 
\[
\frac{a_{\mm{history\, force}}}{a_{\mm{Stokes\, drag}}}\approx\alpha\frac{r}{\eta}
\]
and for the reduction of the slip velocity by memory 
\[
\frac{\left|\v v-\v u\right|_{\mm{memory}}}{\left|\v v-\v u\right|_{\mm{no\, memory}}}\approx\frac{1}{1+\alpha\frac{r}{\eta}}.
\]
The constant $\alpha$ is expected to be on the order of unity and
from numerical simulations we have determined $\alpha=0.69$. These
relations make it possible to quickly obtain an estimate of the importance
of the history force (when $\alpha$ is not known, setting $\alpha=1$
should still yield a reasonable estimate). The fact that these relations
fit well to numerical simulations shows that for the magnitude of
the history force and for the reduction of the slip velocity the determining
parameter is the particle size and that the density is of minor importance. 

This, however, is not true for all effects of memory. We have found
the influence of the history force on particle acceleration, preferential
concentration and collision rates to depend on\emph{ both} the particle
size and density. In these cases we were not able to find simple estimates
as above. A general summary of the findings on preferential concentration
and collision rates is that for a fixed Stokes number a decreasing
density or equivalently an increasing particle size lead to stronger
effects of the history force. These effects can become quite strong
for small densities/large sizes.

An important field of study for inertial particles is the motion of
water droplets in air (e.g. in a cloud), where the density ratio $\varrho\approx1000$
is very large. In this case we have found the effect of memory to
be rather small. This is also true for the relative magnitude of the
the history force and the reduction of the slip velocity, which do
not depend on the particle's density (at least approximately). At
first this might seem contradictory but becomes clear when we note
the following: we considered a fixed range of Stokes numbers ($St\in\left[0.1,3\right]$),
the corresponding particle sizes for $\varrho=1000$ are quite small
($r/\eta\in\left[0.02,0.12\right]$) and hence also the effect of
memory. It might, however, become more significant for larger $r/\eta$.
Even within the investigated range, an interesting finding is that
the change of the collision rate caused by the history force at $\varrho=1000$
and $St=1$ is 8\%. Although this is still small, it is significantly
larger than the effect on all other studied quantities for $\varrho=1000$.
It remains to be seen whether this has any practical implications.\\

Valuable discussions with T. Tél, M. Wilczek, L. Brandt and J. Bec
are acknowledged, as well as the support by the Studienstiftung des
deutschen Volkes. The Eulerian part of the simulation code has been
written by M. Wilczek.

\appendix

\section{Faxén corrections in comparison to the history force\label{sec:faxen-corrections}}

The aim of the present section is to compare the magnitudes of the
Faxén corrections and the history force. We will see that the former
is \emph{much} smaller than the latter for most particle parameters
considered in this paper. 

With the Faxén corrections the evolution equation (\ref{eq:MR-dimensionless})
becomes
\begin{equation}
\dt{\v v}=\beta\Dt{}\left(\v u+\frac{r^{2}}{10}\Delta\v u\right)-\frac{1}{\tau_{p}}\left(\v v-\v u-\frac{r^{2}}{6}\Delta\v u\right)-\sqrt{\frac{3\beta}{\tau_{p}}}\left(\dt{}\right)^{1/2}\left(\v v-\v u-\frac{r^{2}}{6}\Delta\v u\right),\label{eq:MR-dimensionless-faxen}
\end{equation}
see \citet{Gatignol1983,Maxey1983}. The full Faxén corrections are
actually defined in terms of surface and volume integrals of $\v u$
and $\mm D\v u/\mm Dt$. The $\Delta\v u$ terms in (\ref{eq:MR-dimensionless-faxen}),
which are the commonly used forms of Faxén corrections, are the first
non-vanishing terms of a Taylor expansion of these integrals. We restrict
ourself here to these approximations (the full Faxén corrections have
been considered by \citet{Calzavarini2012} in the case of neutrally
buoyant particles through a special numerical scheme). The second
limitation of our comparison is that we do not solve (\ref{eq:MR-dimensionless-faxen})
but equation (\ref{eq:MR-dimensionless}) without Faxén corrections
and only afterwards evaluate $\Delta\v u$ along the particle trajectories.
Our aim is to obtain a rough comparison of the typical magnitudes
of the Faxén corrections and the history force; we expect that our
approach, in spite of these limitations, is sufficient for this purpose.

\begin{figure}
\centering{}\includegraphics{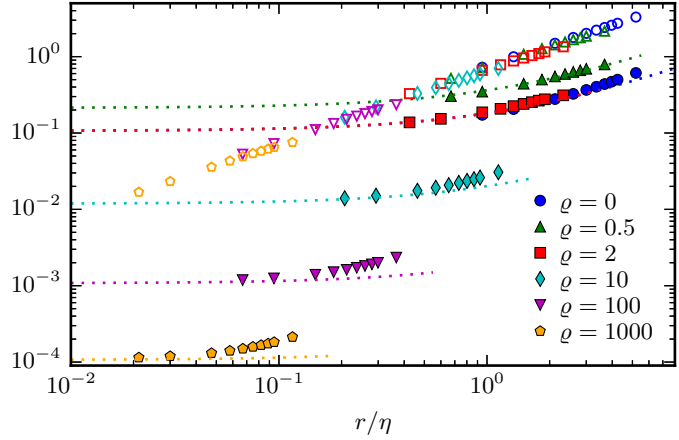}\protect\caption{\label{fig:faxen-ratio}A comparison of the magnitudes of the Faxén
corrections and the history force. Shown are the ratio $\frac{r^{2}}{6}\left\langle \left|\Delta\protect\v u\right|\right\rangle /\left\langle \left|\protect\v v-\protect\v u\right|\right\rangle $
(filled symbols), its approximation (\ref{eq:faxen-ratio-approximation})
(dotted lines) and the ratio $\left\langle a_{H}\right\rangle /\left\langle a_{S}\right\rangle $
(unfilled symbols).}
\end{figure}

In section~\ref{sec:Forces} we have compared the history force to
the drag and found a simple approximate rule for their ratio (\ref{eq:HS-ratio}).
We proceed in the same way for the Faxén corrections, by considering
the ratio to the Stokes drag. The last two terms in (\ref{eq:MR-dimensionless-faxen})
yield the ratio $\frac{r^{2}}{6}\left|\Delta\v u\right|/\left|\v v-\v u\right|$.
By approximating the slip velocity with 
\begin{equation}
\v v-\v u\approx\tau_{p}\left(\beta-1\right)\frac{1}{1+\alpha\frac{r}{\eta}}\Dt{\v u},\label{eq:slip-velocity-estimate-with-history}
\end{equation}
which follows from (\ref{eq:slip-velocity-approximation}) with $\dt{\v v}\approx\Dt{\v u}$,
and using (\ref{eq:tp-definition}) we obtain
\begin{equation}
\frac{r^{2}}{6}\frac{\left|\Delta\v u\right|}{\left|\v v-\v u\right|}\approx\frac{1}{2}\frac{\beta}{\left|\beta-1\right|}\left(1+\alpha\frac{r}{\eta}\right)\frac{\left|\nu\Delta\v u\right|}{\left|\mm D\v u/\mm Dt\right|}.\label{eq:faxen-ratio-approximation}
\end{equation}
Approximation (\ref{eq:slip-velocity-estimate-with-history}) is similar
to (\ref{eq:weakly-inertial-limit}), but is of the slightly higher
order 3/2 in $\tau_{p}$ (as $1/\left(1+\alpha r/\eta\right)\approx1-\alpha r/\eta$
and $r\propto\sqrt{\tau_{p}}$). By averaging over tracers in our
simulation we find
\begin{equation}
\frac{\left\langle \left|\nu\Delta\v u\right|\right\rangle }{\left\langle \left|\mm D\v u/\mm Dt\right|\right\rangle }=0.14.\label{eq:numerical-value}
\end{equation}
Because an average over tracers is equivalent to an Eulerian average,
this ratio can be considered a property of the underlying flow, independent
of the particle picture. Also, this results implies that the main
contribution to $\mm D\v u/\mm Dt=-\nabla p+\nu\Delta\v u$ is from
$\nabla p$ and thus justifies the commonly used name ``pressure
gradient'' for $\mm D\v u/\mm Dt$. Figure~\ref{fig:faxen-ratio}
shows the ratio $\frac{r^{2}}{6}\left\langle \left|\Delta\v u\right|\right\rangle /\left\langle \left|\v v-\v u\right|\right\rangle $
obtained from simulations together with the approximation (\ref{eq:faxen-ratio-approximation})
for the numerical value (\ref{eq:numerical-value}). It fits quite
well for $\varrho<10$. For $\varrho\geq10$ there is a notable mismatch,
nevertheless (\ref{eq:faxen-ratio-approximation}) provides a reasonable
order of magnitude estimate. 

Figure~\ref{fig:faxen-ratio} also shows the ratio of the history
force and the Stokes drag $\left\langle a_{H}\right\rangle /\left\langle a_{S}\right\rangle $,
which was studied in section~\ref{sec:Forces}. We see that for all
studied parameter values the magnitude of the Faxén corrections is
smaller than the magnitude of the history force; in most cases it
is \emph{much} smaller. From all the considered densities the relative
size of the Faxén corrections are largest at $\varrho=0.5$. This
is expected because this value is closest to $\varrho=1$, in which
case $\v v\approx\v u$ and thus the ratio $\frac{r^{2}}{6}\left|\Delta\v u\right|/\left|\v v-\v u\right|$
becomes very large. 

For for $\varrho=1$ we can expect $a_{H}\approx0$ (see also appendix~\ref{sec:Neutrally-buoyant-particles})
and because the Faxén corrections remain finite, they are expected
to be much more important than the history force for neutrally buoyant
particles (indeed even more important than the Stokes drag). 

If we use the approximation (\ref{eq:faxen-ratio-approximation})
and extrapolate to smaller particles sizes, we see that the Faxén
corrections become larger than the history force when $r$ is small
enough. Indeed the ratio $\frac{r^{2}}{6}\left|\Delta\v u\right|/\left|\v v-\v u\right|$
tends to a constant for $r\rightarrow0$ while $a_{H}/a_{S}$ goes
to zero, see figure~\ref{fig:faxen-ratio}. Thus for small $r$ (how
small, depends on $\varrho$) the Faxén corrections become more important
than the history force. However this is not the case for all of our
studied parameter values. Also, when $\varrho$ is not close to $1$
both terms are quite small when the Faxén terms become more important;
then both might be neglected.

Let us now come to the Faxén correction from the first term in (\ref{eq:MR-dimensionless-faxen}).
Using (\ref{eq:slip-velocity-estimate-with-history}) we find the
approximation
\begin{equation}
\frac{\beta\Dt{}\frac{r^{2}}{10}\Delta\v u}{\frac{1}{\tau_{p}}\left|\v v-\v u\right|}\approx\frac{1}{10}\frac{\beta}{\left|\beta-1\right|}\left(\frac{r}{\eta}\right)^{2}\left(1+\alpha\frac{r}{\eta}\right)\frac{\left|\tau_{\eta}\Dt{}\nu\Delta\v u\right|}{\left|\mm D\v u/\mm Dt\right|}\label{eq:faxen-ratio-2}
\end{equation}
and for the unknown ratio 
\[
\frac{\left\langle \left|\tau_{\eta}\Dt{}\nu\Delta\v u\right|\right\rangle }{\left\langle \left|\mm D\v u/\mm Dt\right|\right\rangle }=0.11
\]
from simulations. Using this value and $\alpha=0.69$ (obtained in
section~\ref{sec:Forces}) we can estimate that the Faxén correction
from the first term in (\ref{eq:MR-dimensionless-faxen}) is smaller
than that form the second or third term when $r/\eta\leq2.5$. Furthermore
we find that the ratio (\ref{eq:faxen-ratio-2}) is smaller than $\left\langle a_{H}\right\rangle /\left\langle a_{S}\right\rangle $
for all the considered parameter values (i.e. those shown in figure~\ref{fig:faxen-ratio}).

We thus conclude that the magnitude of the Faxén corrections is smaller
than the magnitude of the history force for all the investigated particle
parameters; in most cases it is much smaller.

\section{Note on neutrally buoyant particles\label{sec:Neutrally-buoyant-particles}}

Up to now we did not consider the case of neutrally buoyant particles.
This case is special because trajectories of tracers, i.e. $\v v=\v u$,
are valid solutions of (\ref{eq:MR-dimensionless}) for $\varrho=1$.
Indeed we have found in our simulations (for Stokes numbers $0.01$,
$0.1$ and $1$) that neutrally buoyant particles behave very similar
to tracers: the mean slip velocity is practically zero, the acceleration
PDF is the same as for tracers, there is no preferential concentration
and collision rates equal that of tracers (with an artificially assigned,
corresponding radius). The inclusion of the history force does not
change any of these findings and thus seems not to be important for
neutrally buoyant particles.

We note that for larger particle sizes (not considered here) neutrally
buoyant particles can behave differently from tracers; \citet{Calzavarini2012}
have studied the effect of memory in this case (along with the influence
of Faxén corrections and non-linear drag).

Concerning the forces acting on a particle, we find $\left\langle a_{S}\right\rangle \approx0$
and $\left\langle a_{H}\right\rangle \approx0$ for all studied Stokes
numbers (see also figure~\ref{fig:olivieri-rho1}a). Thus, the overwhelming
contribution to the particle acceleration comes from the pressure
gradient $a_{P}$. This does not mean that the Stokes drag is unimportant,
because neglecting it (and the history force) in (\ref{eq:MR-dimensionless})
would allow for solutions of the type $\v v=\v u+\v c$ with an arbitrary
constant $\v c$, which is undesirable. One needs the Stokes drag
as a restoring effect which keeps the particle close to a tracer trajectory
(even though it has to act rarely). The history force can be also
considered as a restoring effect (indeed it is sometimes called the
unsteady drag). The fact that we do not find any influence of memory
suggests that the Stokes drag is already sufficient to keep the particles
very close to the tracer trajectories.

Note that we neglected the Faxén corrections in our simulations. This
might be a serious limitation for neutrally buoyant particles as the
findings of appendix~\ref{sec:faxen-corrections} suggest. Thus the
results of this section should to be treated with some caution.

A fundamental question concerning neutrally buoyant particles is whether
the tracer solutions are stable, see e.g. \citet{Babiano2000,Haller2008b}.
The history force might play a very non-trivial role here. For example
it was shown by \citet{Daitche2011} that the history force can destroy
attractors and that the convergence towards attractors becomes very
slow ($\sim t^{-1/2}$) with memory. Thus a thorough study of the
influence of memory on neutrally buoyant particles would have to include
a stability analysis of tracer trajectories; this, however, is beyond
the scope of this paper.

\section{A time-stepping scheme for particles with memory\label{sec:Time-stepping-scheme-appendix}}

Let us start by rewriting the evolution equation (\ref{eq:MR-dimensionless})
for the slip velocity $\v w=\v v-\v u$:
\begin{align}
\dt{\v w} & =\left(\beta-1\right)\dt{\v u}-\beta\v w\cdot\nabla\v u-\frac{1}{\tau_{p}}\v w-\sqrt{\frac{3\beta}{\pi\tau_{p}}}\dt{}\int_{0}^{t}\frac{\v w}{\sqrt{t-\tau}}\,\mm d\tau\label{eq:MR-slipvelocity}\\
\dt{\v x} & =\v w+\v u,
\end{align}
where $\mm d\v u/\mm dt=\partial_{t}\v u+\v v\cdot\nabla\v u$. We
wish to obtain a numerical solution at the time points $t_{n}=nh$,
where $h$ is the time step. Integrating (\ref{eq:MR-slipvelocity})
from $t_{n}$ to $t_{n+1}$ we get
\begin{multline}
\v w_{n+1}=\v w_{n}+\left(\beta-1\right)\left(\v u_{n+1}-\v u_{n}\right)-\beta\int_{t_{n}}^{t_{n+1}}\v w\cdot\nabla\v u\,\mm d\tau-\frac{1}{\tau_{p}}\int_{t_{n}}^{t_{n+1}}\v w\,\mm d\tau\\
+\xi\int_{0}^{t_{n+1}}\frac{\v w}{\sqrt{t_{n+1}-\tau}}\,\mm d\tau-\xi\int_{0}^{t_{n}}\frac{\v w}{\sqrt{t_{n}-\tau}}\,\mm d\tau\label{eq:MR-integrated}
\end{multline}
where $\xi=-\sqrt{3\beta/(\pi\tau_{p})}$. To obtain a numerical stepping
scheme we have to approximate the integrals. We begin with the first
two:
\begin{equation}
\int_{t_{n}}^{t_{n+1}}f(\tau)\mm d\tau\approx h\sum_{i=0}^{m}\lambda_{i}^{(m)}f(\tau_{n+1-i})
\end{equation}
The coefficients $\lambda_{i}^{(m)}$ are well known. For an explicit
scheme, i.e. $\lambda_{0}=0$, they are the coefficients $\v{\lambda}^{B,m}$
of the Adams-Bashforth schemes:
\begin{eqnarray}
\v{\lambda}^{B,2} & = & \left(0,1\right)\\
\v{\lambda}^{B,3} & = & \left(0,\frac{3}{2},-\frac{1}{2}\right)\\
\v{\lambda}^{B,4} & = & \left(0,\frac{23}{12},-\frac{4}{3},\frac{5}{12}\right)
\end{eqnarray}
For an implicit scheme they are the coefficients $\v{\lambda}^{M,m}$
of the Adams-Moulton schemes: 
\begin{eqnarray}
\v{\lambda}^{M,2} & = & \left(\frac{1}{2},\frac{1}{2}\right)\\
\v{\lambda}^{M,3} & = & \left(\frac{5}{12},\frac{8}{12},-\frac{1}{12}\right)
\end{eqnarray}
The errors of these approximations are $O(h^{m})$ for $\v{\lambda}^{B,m}$
and $O(h^{m+1})$ for $\v{\lambda}^{M,m}$. The history integrals
in (\ref{eq:MR-integrated}) are approximated with 
\begin{equation}
\int_{0}^{t_{n}}\frac{\v w}{\sqrt{t-\tau}}\mm d\tau\approx\sqrt{h}\sum_{i=0}^{n}\mu_{i}\v w_{n-i}
\end{equation}
where $\v w_{n}=\v w(t_{n})$ and the coefficients $\mu_{i}$ are
given by \citet{Daitche2013}. We thus obtain the following scheme
for the slip velocity:
\begin{multline}
\v w_{n+1}=\v w_{n}+\left(\beta-1\right)\left(\v u_{n+1}-\v u_{n}\right)-\beta h\sum_{i=0}^{m}\lambda_{i}^{B,m}\left[\v w\cdot\nabla\v u\right]_{n+1-i}-\frac{h}{\tau_{p}}\sum_{i=0}^{m}\lambda_{i}^{M,m}\v w_{n+1-i}\\
+\xi\sqrt{h}\left(\sum_{i=0}^{n+1}\mu_{i}\v w_{n+1-i}-\sum_{i=0}^{n}\mu_{i}\v w_{n-i}\right)
\end{multline}
Note that the right-hand side contains $\v w_{n+1}$, but only linearly.
Bringing it to the left-hand side yields an explicit scheme:
\begin{eqnarray}
\left(1+\frac{h}{\tau_{p}}\lambda_{0}^{M}-\xi\sqrt{h}\mu_{0}\right)\v w_{n+1} & = & \v w_{n}+\left(\beta-1\right)\left(\v u_{n+1}-\v u_{n}\right)\nonumber \\
 &  & -\beta h\sum_{i=0}^{m}\lambda_{i}^{B,m}\left[\v w\cdot\nabla\v u\right]_{n+1-i}-\frac{h}{\tau_{p}}\sum_{i=1}^{m}\lambda_{i}^{M,m}\v w_{n+1-i}\nonumber \\
 &  & +\xi\sqrt{h}\left(\sum_{i=1}^{n+1}\mu_{i}\v w_{n+1-i}-\sum_{i=0}^{n}\mu_{i}\v w_{n-i}\right)\label{eq:scheme-1}\\
\v x_{n+1} & = & \v x_{n}+h\sum_{i=0}^{m}\lambda_{i}^{B,m}\left(\v w_{n+1-i}+\v u_{n+1-i}\right)\label{eq:scheme-2}
\end{eqnarray}
Using $\lambda_{i}^{B,4}$, $\lambda_{i}^{M,3}$ and the third-order
coefficients $\mu_{i}$ from (\citet{Daitche2013}) the above scheme
has a one-step error of $O(h^{4})$, i.e. it is a third-order scheme.

As this scheme relies on previous values of $\v x$ and $\v w$, one
needs to start the integration with lower order schemes: the first
step with $\lambda_{i}^{B,2}$, $\lambda_{i}^{M,1}$ and the first-order
$\mu_{i}$; the second step with $\lambda_{i}^{B,3}$, $\lambda_{i}^{M,2}$
and the second-order $\mu_{i}$.

To further improve the accuracy of the scheme, corrector steps have
been applied. A corrector step is performed by predicting $\v x_{n+1}$,
$\v w_{n+1}$ with the explicit scheme (\ref{eq:scheme-1})-(\ref{eq:scheme-2}),
replacing the $\lambda_{i}^{B}$ in this scheme with $\lambda_{i}^{M}$
(thus making it implicit), inserting the predicted $\v x_{n+1}$,
$\v w_{n+1}$ in the right-hand side and calculating a corrected estimate
for $\v x_{n+1}$, $\v w_{n+1}$. (The corrector steps are not essential
for the numerical scheme.)

That the one-step error of this scheme scales indeed with $h^{4}$,
has been verified in a case where an analytical solution of (\ref{eq:MR-dimensionless})
is available: a particle moving in a two dimensional vortex (\citet{Angilella2004}),
see also (\citet{Daitche2013}).

\section{Forcing the turbulence\label{sec:Forcing}}

We solve the vorticity equation in a triply periodic box of length
$L_{\mm{box}}=2\pi$. In Fourier space the equation becomes 
\[
\partial_{t}\v{\omega}_{\v k}=i\v k\times\mathcal{F}[\v u\times\v{\omega}]_{\v k}-\nu\v k^{2}\v{\omega}_{\v k}+\v F_{\v k}.
\]
Here $\v{\omega}=\nabla\times\v u$ is the vorticity, $\v k\in\mathbb{Z}^{3}$
the wave vector, $\mathcal{F}$ the Fourier transform and $\v F$
the forcing. The latter is implicitly defined as described in the
following. We force by inserting the energy lost during one time step
back into large scale modes $\v{\omega}_{\v k}$ with $\v k$ in the
forcing band 
\[
B=\left\{ \v k\left|3\leq\v k^{2}\leq9\right.\right\} .
\]
Let $\v{\omega}_{\v k}^{n}$ be the current vorticity and $\tilde{\v{\omega}}_{\v k}^{n+1}$
the vorticity after one time step without forcing (more precisely,
when using a scheme consisting of multiple stages, like a Runge-Kutta
method, $\tilde{\v{\omega}}_{\v k}^{n+1}$ is the vorticity after
one \emph{stage}). The forcing is applied by rescaling the modes in
the forcing band: 
\[
\v{\omega}_{\v k}^{n+1}=\begin{cases}
f\tilde{\v{\omega}}_{\v k}^{n+1} & \v k\in B\\
\tilde{\v{\omega}}_{\v k}^{n+1} & \v k\notin B
\end{cases}
\]
The factor $f$ is chosen such that the energy of the flow 
\[
E\left[\v{\omega}_{\v k}\right]=\frac{1}{2}\sum_{\v k}\frac{\left|\v{\omega}_{\v k}\right|^{2}}{\v k^{2}}
\]
remains constant, i.e. 
\[
E\left[\v{\omega}_{\v k}^{n+1}\right]=E\left[\v{\omega}_{\v k}^{n}\right].
\]
The modes$\left\{ \left.\v{\omega}_{\v k}\right|\v k\notin B\right\} $
evolve freely and the energy is kept constant by inserting energy
into the modes $\left\{ \left.\v{\omega}_{\v k}\right|\v k\in B\right\} $.
Note also that only the modulus of the modes is modified, their direction
can evolve freely, which improves the statistical isotropy of the
flow. 

\bibliographystyle{jfm}
\bibliography{inertial_particles}

\end{document}